\documentclass[preprint,amsmath,amssymb,prX,nofootinbib]{revtex4}
\pdfoutput=1

\usepackage{graphicx}
\usepackage{dcolumn}
\usepackage{bm}

\begin{document}


\title{Topological characterization of flow structures in resistive pressure-gradient-driven turbulence}

\author{B. A. Carreras}
 \affiliation{BACV Solutions Inc., Oak Ridge, Tennessee 37830, U.S.A.}
\email{bacv@comcast.net}

\author{I. Llerena}
 \affiliation{Department d'\`Algebra i Geometria, Facultat de Matem\`atiques, Universitat de Barcelona, Barcelona, Spain}

\author{L. Garcia}
 \affiliation{Departamento de F{\'\i}sica, Universidad Carlos III de Madrid, 28911 Legan\'es, Madrid, Spain}

\author{I. Calvo}
 \affiliation{Laboratorio Nacional de Fusi\'on, Asociaci\'on EURATOM-CIEMAT, 28040 Madrid, Spain}

\date{\today}

\begin{abstract}

Visualization of turbulent flows is a powerful tool to help understand the turbulence dynamics and induced transport. However, it does not provide a quantitative description of the observed structures. In this paper, an approach to characterize quantitatively the topology of the flows is given. The technique, which can be applied to any type of turbulence dynamics, is illustrated through the example of resistive ballooning instabilities.

\end{abstract}

\maketitle

\section{INTRODUCTION}
\label{sec:introduction}

Visualization of flow structures plays an important role in providing understanding of the turbulence dynamics and some of the properties of the transport mechanisms induced by the same turbulence. In the case of turbulent plasmas in toroidal geometry, flow structures may be quite complicated \cite{Carreras_03}.  Eddies form toroidal knots, which through ballooning effects can merge in the outer regions of the torus, forming what has been called streamers \cite{Beyer_00}.  In the inner toroidal region, there is a filamentation of these eddies and the resulting flow filaments twist around the torus following the magnetic field lines. When these ballooning structures remain in the plasma in quasi-steady state, they can induce pseudo-chaotic behavior and anomalous diffusion of tracer particles both in the radial and poloidal directions \cite{Zaslavsky_05, Calvo_08}.

\begin{figure}
\centering
\includegraphics[angle=0,width=15cm]{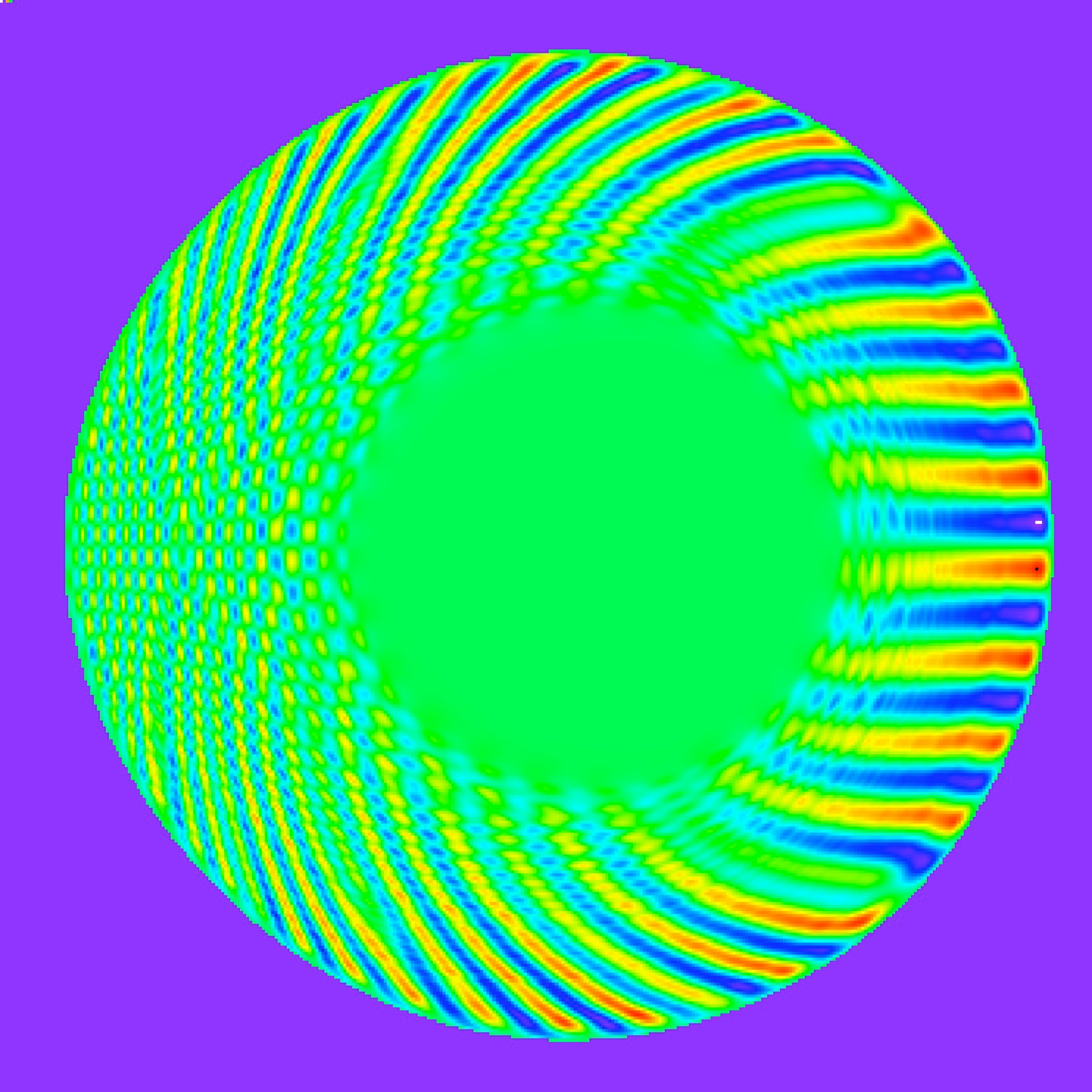}
\caption{Contours of $\Phi={\rm{Constant}}$ on the toroidal cross
section with $\zeta=0$. Streamers form in the outer region (right) of
the torus.}
\label{FIG:1}
\end{figure}

Although visualization is a powerful tool in interpreting numerical results, it does not provide by itself a quantitative characterization of the flow structures.  Here, we take a step towards this quantitative characterization by introducing diagnostics that measure some topological properties of the structures. The introduction of the proposed techniques will be performed through their application to a concrete physical problem: the study of the topology of the flows in resistive pressure-gradient-driven turbulence in toroidal plasmas. We use the set of reduced magnetohydrodynamic (MHD) equations presented in Refs.~\cite{Strauss_77, Drake_84}, so that the flow is determined by the velocity stream function $\Phi \left( {\rho ,\theta ,\zeta } \right)$ where $( {\rho ,\theta ,\zeta } )$ are toroidal coordinates. The velocity field, ${\bf{V}}$, is given in terms of the stream function as ${\bf{V}}=\nabla \Phi \times {\bf{b}}$, where ${\bf{b}}={\bf{B}}/\left|{{\bf{B}}}\right|$ is a unit vector in the direction of the magnetic field. A 2-D plot of the constant $\Phi$ surfaces from a low-$\beta$ resistive pressure-gradient-driven turbulence calculation is given in Fig.~\ref{FIG:1}, showing the aforementioned streamers in the outer region and the filamentation in the inner region. Algebraic topology can be used to measure the complexity of these structures. 

\begin{figure}
\centering
\includegraphics[angle=0,width=15cm]{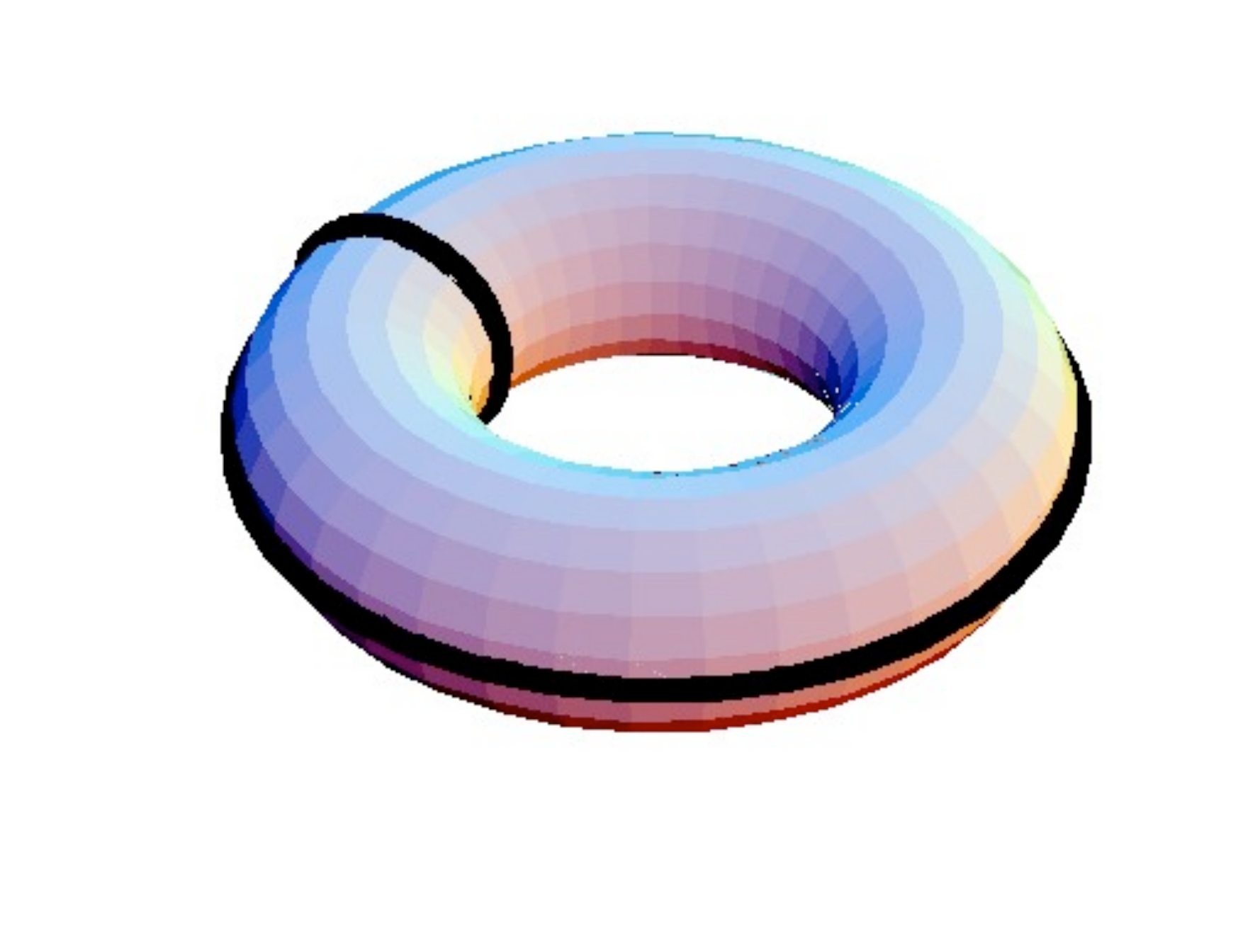}
\caption{Toroidal surface showing two independent cycles that are not
contractile to a point.}
\label{FIG:2}
\end{figure}

For structures in ${\mathbb{R}}^3$ there are three relevant topological invariants (i.e. quantities which do not change under continuous deformations of the structure): the first three Betti numbers \cite{Munkres_84} that we denote by $b_0$, $b_1$ and $b_2$. $b_0$ is the number of arc-connected components of the structure, $b_1$ is the number of (independent) non-contractible loops, and $b_2$ is the number of voids. For instance, the solid sphere has $b_0 = 1$, $b_1 = 0$ and $b_2 = 0$, because it has one connected component, no voids and any loop can be contracted to a point. However, the sphere has $b_0 = 1$, $b_1 = 0$ and $b_2 = 1$, because it has a void. The toroidal surface has two independent non-trivial loops (see Fig.~\ref{FIG:2}) and one void, so that $b_0 = 1$, $b_1 = 2$ and $b_2 = 1$, but for a solid torus the Betti numbers are $b_0 = 1$, $b_1 = 1$ and $b_2 = 0$ because it only has one non-contractible loop.

A crucial point is that the Betti numbers are the rank of commutative groups obtained from a succession of morphisms between free commutative groups; they make up the {\em{chain complex}} of the space. Furthermore, the Betti numbers also are the dimension of vector spaces obtained from a succession of linear maps. As a result, linear algebra methods are used to compute them. 

For the complicated flow structures obtained in numerical calculations the determination of the Betti numbers is far from trivial. The classical algorithms for the computation of the homology of a set \cite{Munkres_84} (the Betti numbers are the ranks of the homology groups) yield exponential run time. Many efforts have been done to reduce the run time and some rather efficient software has been developed.  Here, we use the software package provided by the Computational Homology Project (CHomP) \cite{Kaczynski_00, Kaczynski_04} in order to diagnose the numerical results from 3-D plasma turbulence calculations. The CHomP project has developed many numerical tools. A description of such tools and some examples of applications can be found in the CHomP website \footnotemark.

There are several ways to extract the information needed to construct a chain complex of a set in $\mathbb R^d$.  CHomP uses a decomposition of the set in cubes, i.e. it uses a union of cubes that approximates our set as finely as possible. The input data of CHomP are either text files listing all those cubes, or a bitmap with each pixel representing a cube. Ref.~\cite{Kaczynski_04} gives a systematic computational approach to the homology of cubical sets, which is the approach that we follow here.

\footnotetext{http://chomp.rutgers.edu/}

One of the most serious issues that we had to face when calculating the Betti numbers with the CHomP software was to determine the right resolution of the 3-D structures, since we easily reach the capacity limit of the software. To understand the problems involved in finding the proper resolution, we have first investigated, with appropriate analytical models, the most efficient way of providing data about the 3-D structures to the CHomP software in order to have converged results for the Betti numbers. This is discussed in Section \ref{sec:calculation}. In Section \ref{sec:alternative}, we present an alternative, somehow indirect method for deducing the Betti numbers based on 2-D computations with CHomP. In this way we can test the numerical convergence and have a verification of the results of Section \ref{sec:calculation}. For simple situations of ballooning structures, we give in Section \ref{sec:analytical} an analytical calculation of the Betti numbers. This provides a way to validate the numerical approach.

In this paper, we limit the analysis and calculations to a single toroidal ballooning structure obtained either through an analytical parameterization or a numerical calculation. Studies are under way to calculate the Betti numbers during the time evolution of the resistive pressure-gradient-drive turbulence for different beta values.

\section{OPTIMIZATION OF THE CUBE-COVERING OF THE FLOW}
\label{sec:calculation}

As mentioned in the Introduction, we present our approach to the quantitative study of flow topology by means of the particular example of pressure-gradient-driven turbulence in toroidal plasmas. The numerical results are the steady state solutions given by the 3D code FAR  \cite{Charlton_86}, which solves the set of reduced MHD equations \cite{Strauss_77, Drake_84} as an initial value problem. In this code, all fields are Fourier expanded in the poloidal and toroidal angles. We denote by $m$ ($N$) the poloidal (toroidal) mode number. A finite difference representation is used for the radial coordinate $\rho$. The numerical scheme is implicit for the linear terms, and explicit for the nonlinear ones. The evolution is implemented through two half-time steps to ensure second-order accuracy in time. Details of the numerical calculations are given in Ref.~\cite{Garcia_02}.

For very low-$\beta$ plasmas, a single toroidal mode, $N$, dominates the spectrum \cite{Garcia_02}. Here, we analyze the topological structure of some of these solutions.  Since the functions are defined on a grid in $\rho$, we either interpolate the values using splines or use a parameterization based on the linear structure of these modes. We also use this parameterization as a simplified model for these structures and in the verification and validation calculations.

For a single toroidal mode $N$, we use a parametrization based on the linear eigenmode structure of the resistive ballooning~\cite{Garcia_99}:

\begin{equation}
\label{eq:single_mode}
\Phi \left( {\rho ,\theta ,\zeta } \right) = A\sum\limits_{m = m_0 }^{m_1 } {\exp \left[ { - \frac{{\left( {m - m_M } \right)^2 }}
{{2W_B^2 }}} \right]} \exp \left[ { - \frac{{\left( {\rho  - \rho _m } \right)^2 }}
{{2W_m^2 }}} \right]\sin \left( {m\theta  + N\zeta } \right).
\end{equation}
Here, $m_M$ is the dominant poloidal mode and $\rho_m$ is the radial location of the singular surface with $q(\rho_m)=m/N$, where $q$ is the safety factor. For a parabolic $q$-profile, $q(\rho)=q(0)+\left[{q(1)-q(0)}\right] \rho^2$, we have

\begin{equation}
\label{eq:rho_m}
\rho _m  = \sqrt {\frac{{{m / N} - q\left( 0 \right)}}
{{q\left( 1 \right) - q\left( 0 \right)}}} .
\end{equation}

Each of the components of the ballooning mode, Eq.~(\ref{eq:single_mode}), is characterized by a width $W_m$. When we use the parameterization (\ref{eq:single_mode}) as a simple model in the covering optimization tests (see below), we usually take $A = 1$, $W_B = \infty$, and $W_m = W$. In this way, for a given value of $N$, the ballooning structure is determined by the single parameter $W$. When we take the FAR numerical results as input for the CHomP software, we either fit the parameters $A$, $W_B$, and $W_m$ to the numerical data or use splines to interpolate between radial points.

The CHomP software computes the Betti numbers of spaces that are union of a finite number of $n$-dimensional cubes, with $n \le 26$; more precisely, spaces that are union of a finite number of cubes obtained from of the unit $n$-cube $[0,1]^n$ by translation with a vector of entire coordinates.  We take $[0,1]^0$ to be the origin. Thus the data files used by the CHomP program are in the form of a list of cubes, which cover the structure. For a structure in $\mathbb R^3$, the maximum number of cubes that can be used is $2.6\times 10^8$. Therefore, when we have a complicated numerical structure it is wise to find the optimal set of cubes that cover it for a fixed number of cubes.

To optimize the cube selection, the first step is to work on the most efficient coordinate system. In the case of magnetically confined plasmas in toroidal geometry, we can use Cartesian coordinates in real space, but they are inefficient because Cartesian cubes do not adapt well to the curvature of the toroidal shape. Toroidal coordinates $( {\rho ,\theta ,\zeta } )$ are better suited. Here, $\rho$ is a radius-like equilibrium flux surface label, $\theta$ is the poloidal angle and $\zeta$ the toroidal angle. The usage of this coordinate system requires choosing the periodicity option for $\theta$ and $\zeta$ in the CHomP code. When we use toroidal coordinates and for a fixed cube size, we need two orders of magnitude less cubes than for Cartesian coordinates with the same cube size. These tests were done for a single $N = 5$ structure.

Most of the numerical tests have been done for toroidal mode numbers in the range $5Ê\leÊNÊ\leÊ23$. The case $NÊ=Ê23$ is motivated by the results of the nonlinear calculations~\cite{Garcia_02}. This mode dominates the toroidal mode spectrum for some of the numerical results. We also use lower values of $N$ to study the scaling of the numerical properties and because for low $N$ the calculations are faster and the representation of the structures is simpler.

The Betti numbers, as any topological invariant, remain unchanged under deformations. Therefore, another way of optimizing the cube selection is stretching or shrinking the coordinate directions, i.e. making a coordinate transformation $( {a\rho ,b\theta ,c\zeta } )$, for three convenient constants $a$, $b$ and $c$. By means of such a change we can choose the number of cubes in each direction that cover our torus.  In practice, we see that we do not need as much resolution, i.e. as many cubes, in the toroidal as in the poloidal and radial directions. It turns out that a good choice is $N_\rho =  N_\theta  =  2N_\zeta$, where  $N_\rho$,  $N_\theta$  and   $N_\zeta$  are the number of cubes in the radial, poloidal and toroidal direction respectively.

Once we have our torus covered with cubes, the next important question is how to prescribe that a cube belongs to the structure. For a cube to be in the structure, we require that a certain number $k$ of its vertices belongs to the structure; in this case we say that the cube is black.

\begin{figure}
\centering
\includegraphics[angle=0,width=15cm]{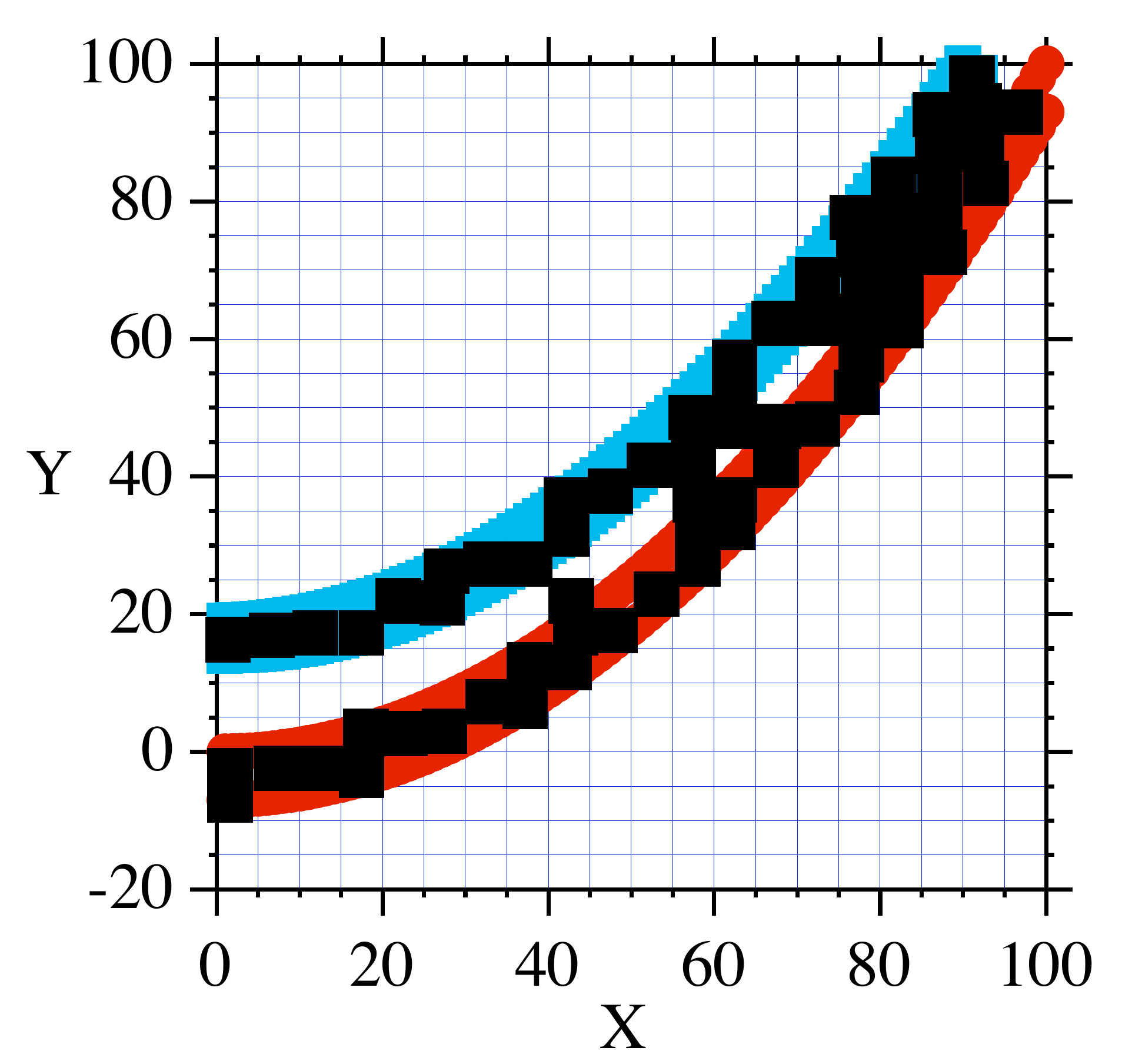}
\caption{2-D representation of the covering by squares of two
filaments running nearly parallels. The regions in red and blue
represent the filaments. The covering cubes are black if the
right-upper vertex is either red or blue. We can see the formation of
false cycles by the black squares (2-D cubes) covering the filaments.}
\label{FIG:3}
\end{figure}

At a fixed time $t$, we define a flow structure as the set of points such that  $\Phi \left( {\rho ,\theta ,\zeta, t } \right) \ge \Phi_0$, where $\Phi_0$ is a constant. A reasonable prescription is that a cube is black if it has $k$ or more vertices verifying the condition $\Phi \left( {\rho ,\theta ,\zeta, t } \right) \ge \Phi_0$. If $k$ is too small there is a problem of creation of false loops (overestimation of $b_1$): since the filaments of the ballooning structure run quasi-parallel for some toroidal distances and close to each other, if many vertices of the cubes lie outside the filaments they can touch and cause the false loops. Fig.~\ref{FIG:3} shows an example of this spurious effect in 2-D. In this figure the filaments are represented by the regions in red and blue, and we have chosen to consider a cube black if the right-upper vertex is either red or blue. We can see the formation of false cycles by the black squares (2-D cubes) covering the filaments.

On the other hand, if $k$ is too large, there are two possible reasons for obtaining false results. Firstly, we can have regions in the structure that are thinner than the cubes, causing false breaks of the structure, and leading to an overestimation of the number of connected components ($b_0$) and an underestimation of the number of loops ($b_1$). Secondly, the existence of filaments very close to each other, with separation of the order of the cube size, produces the formation of false loops for reasons analogous to those shown in Fig.~\ref{FIG:3}. In our studies we found that this second effect on $b_1$ is more important than the first, so we have an overall increase of false loops.

Therefore, in the optimization studies we see that the best selection of $k$ is characterized by giving a minimum in $b_1$ and the start of an increase in $b_0$. This optimal value of $k$ gives the maximum resolution of the structure for a fixed cube size.

\begin{figure}
\centering
\includegraphics[angle=0,width=15cm]{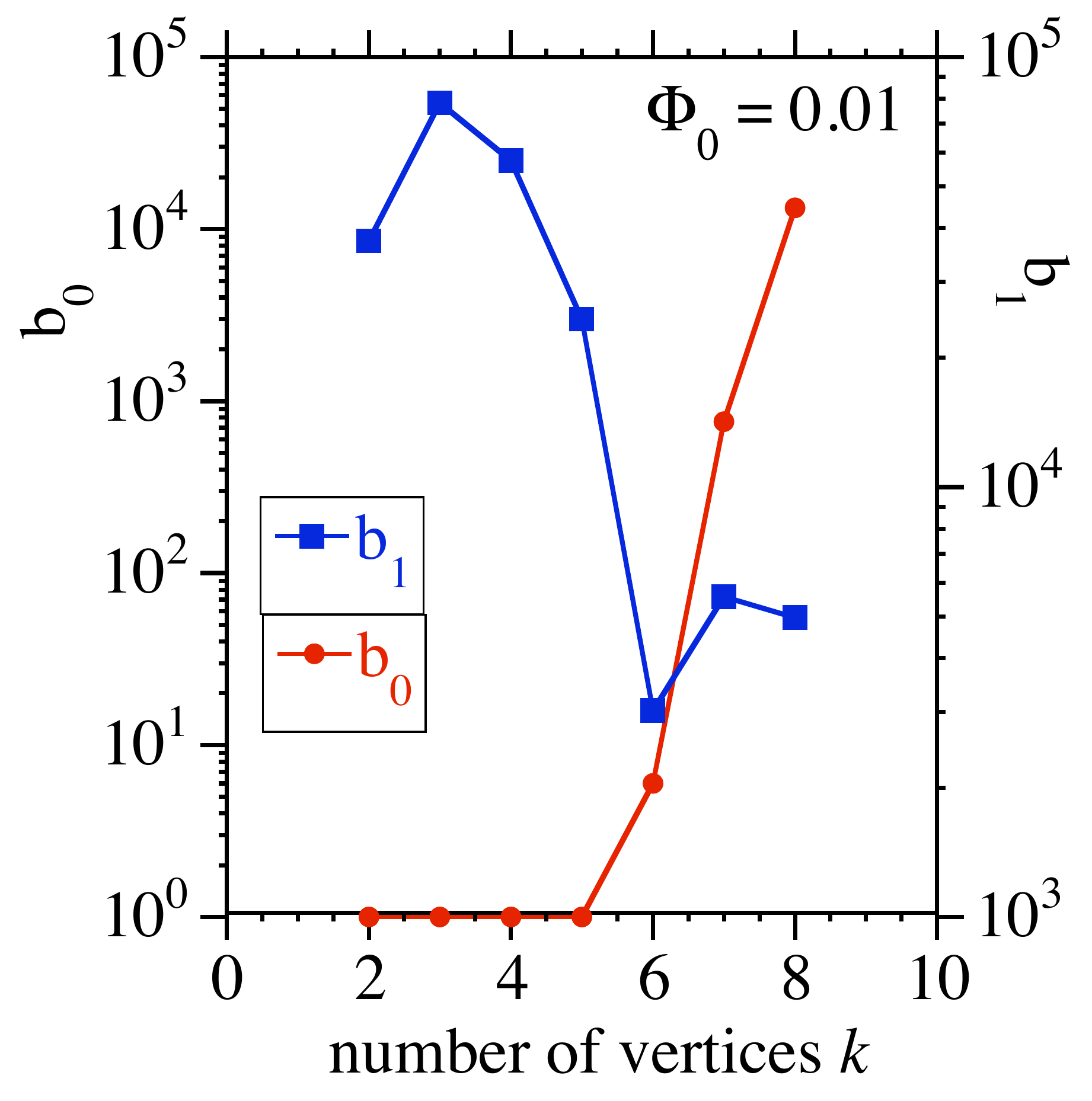}
\caption{Betti numbers $b_0$ and $b_1$ for an $N= 23$ flow structure
as a function of $k$, the number of vertices in the structure that
defines a black cube. The calculation is for a fixed size cubes with
$N_\rho=N_\theta=2N_\zeta=200$.}
\label{FIG:4}
\end{figure}

\begin{figure}
\centering
\includegraphics[angle=0,width=15cm]{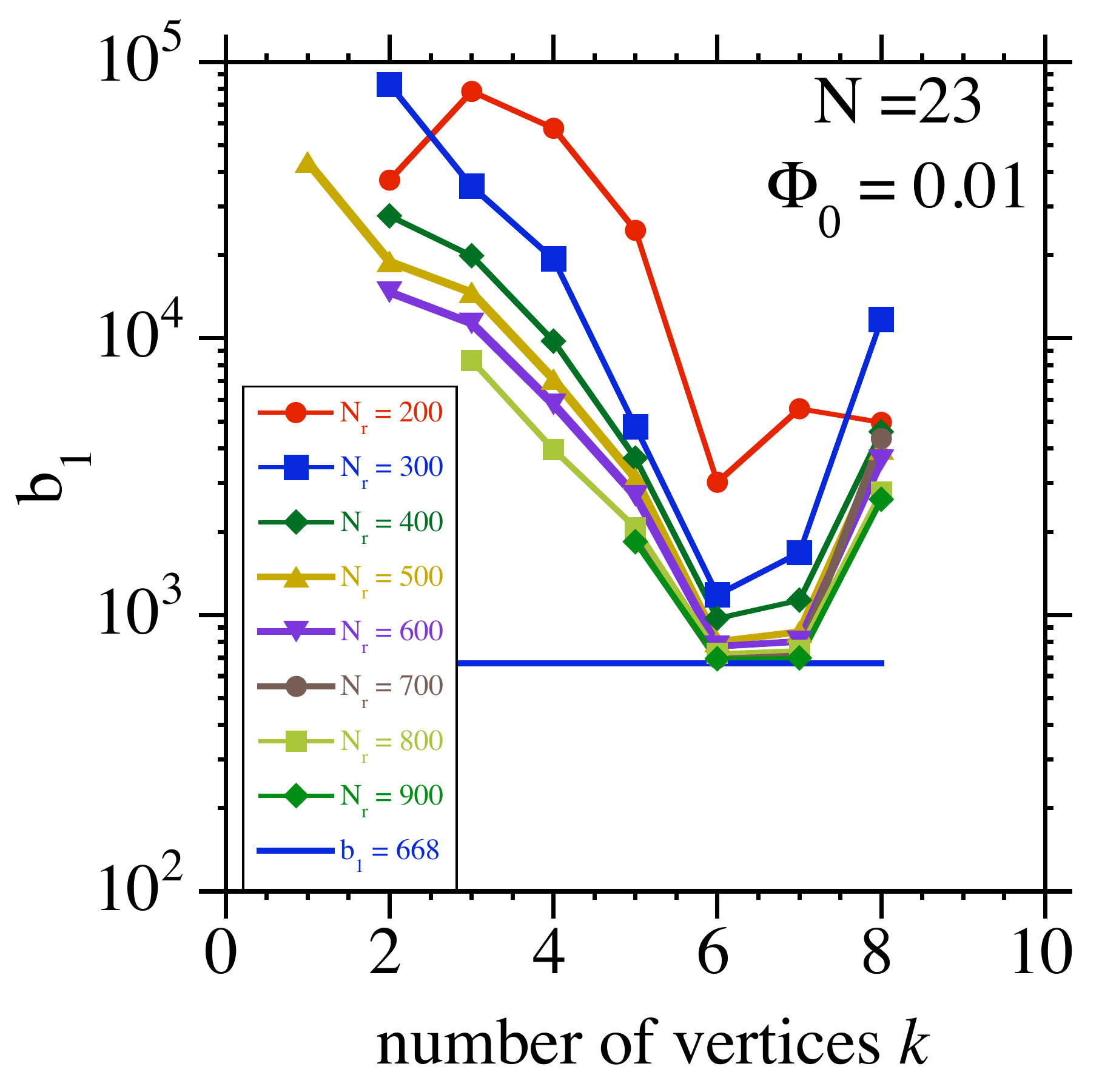}
\caption{Betti number $b_1$ for the same $N= 23$ flow structure as in
Figure~\ref{FIG:4} as a function of $k$, the number of vertices in the
structure that define a black cube. The calculation shows the
convergence as the number of cubes increase for cubes coverings
verifying $N_\rho=N_\theta=2N_\zeta$.}
\label{FIG:5}
\end{figure}

For the parameters in Ref.~\cite{Garcia_02} and $\beta_0= 0.003$, the dominant mode is $N = 23$. We fitted the numerical results for this mode using Eq.~(\ref{eq:single_mode}) and with this parameterization constructed the set of cubes representing the flow structure for $k$ going from 1 to 8. For the flow structure with $\Phi_0=0.01$ and in Fig.~\ref{FIG:4}, we plotted $b_0$ and $b_1$ for fixed size cubes, $N_\rho =  N_\theta  =  2N_\zeta=200$ as a function of $k$. The basic picture is as expected. However, the effect of reduction of $b_1$ is very dramatic as $k$ increases; there is a change of three orders of magnitude in going from $k =2$ to $k = 6$. The optimization works for all the different cube sizes that we have tried. In Fig.~\ref{FIG:5} and for the same flow structure as in Fig.~\ref{FIG:4}, we have plotted the results of a systematic calculation of $b_1$ for different numbers of cubes and for each set of cubes varying the number of vertices $k$ used in defining a black cube. In all cases, we have used $N_\rho =  N_\theta  =  2N_\zeta$. We have also plotted in the same graph the value $b_1 = 668$ which is the same value for this configuration obtained using an alternative approach discussed in Section \ref{sec:analytical}. We see that all values of $b_1$ tend to converge to the expected value; however, convergence is very much faster for the case of $k = 6$ and $k = 7$ vertices. We have obtained similar results for different values of $\Phi_0$.

\section{AN ALTERNATIVE APPROACH TO THE CALCULATION OF THE BETTI NUMBERS OF A FLOW STRUCTURE}
\label{sec:alternative}

Resolution of the numerical calculations is a systematic issue in the determination of the Betti numbers. Therefore, it is important to have different ways to calculate them in order to do validation of the results. An alternative method to the one discussed in the previous section is to consider many toroidal cuts, calculate the 2-D Betti numbers of the toroidal sections of the structures and infer the Betti numbers of the 3-D structure. The advantage of this approach is that in 2-D calculations we can achieve very high resolution. The disadvantage is that it requires performing the 2-D calculations in many toroidal cuts if we do not want to miss some of the features of the structure (and it is impossible to know {\em{a priori}} how many cuts we need).

In 2-D there are only two relevant Betti numbers: the number of connected components, $b_0$, and the number of loops, $b_1$. For the structures we are looking for, and in the case of a single $N$ mode, $b_1$ in 2-D is generally zero and all the 2D components are contractible.

Another important result on Betti numbers is that their alternate sum coincides with the alternate sum of the number of cubes in different dimensions \cite{Munkres_84}. This number is known as the Euler-Poincar\'e characteristic of the set. Assume, for simplicity, that we have a cubical space in $\mathbb R^2$ or $\mathbb R^3$. By cubes of dimension 0 we mean the vertices of the cubes that form the cubical set. By cubes in dimension 1 we mean the edges of these cubes. In $\mathbb R^2$ the maximal cubes have dimension 2 and are squares. In $\mathbb R^3$ the maximal cubes have dimension 3 and their faces are cubes in dimension 2. For example, let $N_v$, $N_e$ and $N_s$ be the number of vertices, edges and squares of the union of a family of squares in $\mathbb R^2$. Then:

\begin{equation}
\label{eq:chi_def}
\chi=b_0-b_1+b_2=N_v-N_e+N_s.
\end{equation}

In a ballooning structure, the filaments can be contracted to a 1-D subspace, which is just a graph, i.e. a structure formed by points, called vertices, and lines joining them, called edges. As this graph is a deformation retract of the filaments, it has the same Betti numbers and the same Euler-Poincar\'e characteristic.			

\begin{figure}
\centering
\includegraphics[angle=0,width=7.2cm]{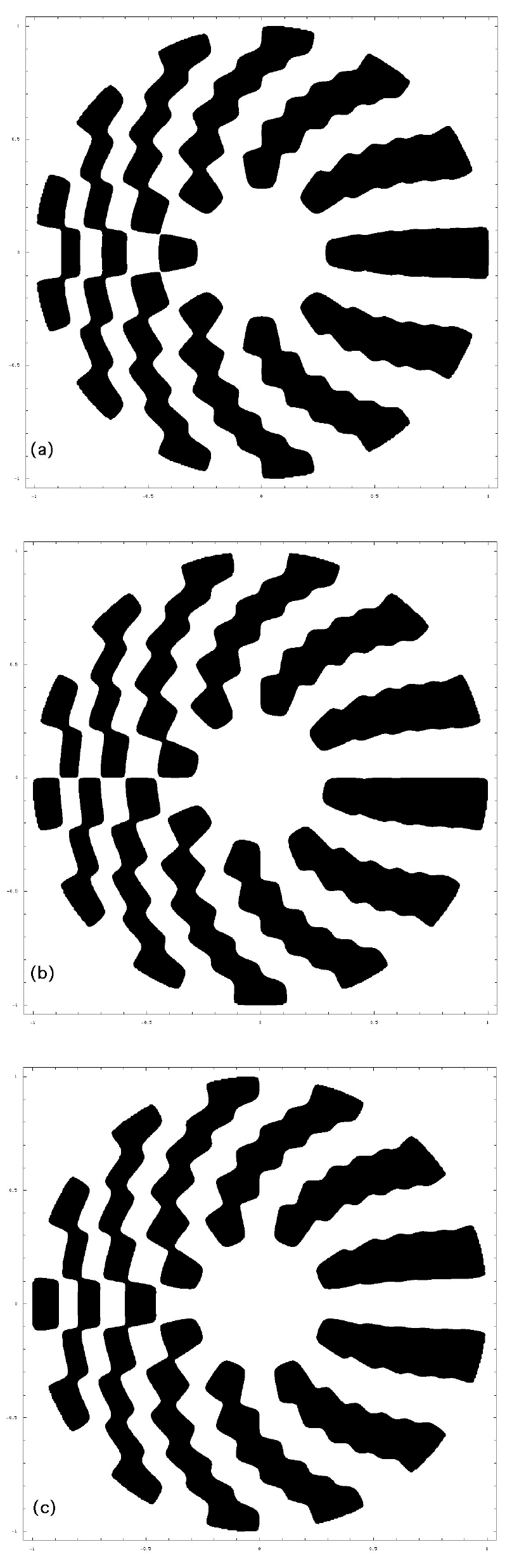}
\caption{Regions with $\Phi \left( {\rho ,\theta ,\zeta = \zeta _0 }
\right) \ge \Phi _0$ for a $N= 7$ flow structure. Here
$\Phi_0=0.01$. We have plotted three values of $\zeta_0$: a)
$\zeta_0=\pi/14$, b) $\zeta_0=\pi/7$, and c) $\zeta_0=3\pi/14$.}
\label{FIG:6}
\end{figure}

\begin{figure}
\centering
\includegraphics[angle=0,width=15cm]{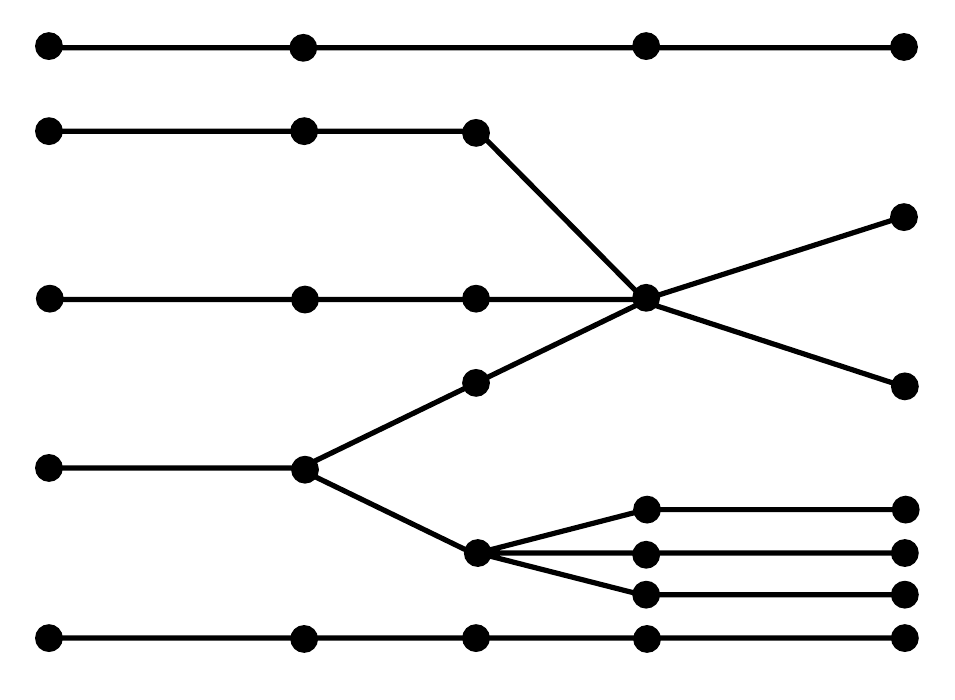}
\caption{Graph representing the changes in the connected components as
we move in the toroidal direction. The nodes represent the connected
components in a given toroidal cut.}
\label{FIG:7}
\end{figure}

Let us consider for instance a toroidal cut of a given $N = 7$ structure. In Fig.~\ref{FIG:6}, we have plotted in black the regions with $\Phi \left( {\rho ,\theta ,\zeta=\zeta_0} \right) \ge \Phi_0$ for $\Phi_0=0.01$ and three values of $\zeta_0$. In Fig.~\ref{FIG:6}a, we can see 10 black connected components. As $\zeta_0$ changes the number of connected component changes; they break and merge as we move in the $\zeta$ direction. In general, given a toroidal cut we will have $E$ well-defined structures. These are identified as the first $E$ vertices of our graph. In Fig.~\ref{FIG:7} we have drawn one such graph with $E = 5$. As we move toroidally, we draw the corresponding number of edges; if we started with $E$ vertices we add $E$ edges. Assume that at some toroidal position a filament breaks up into $j$ components. This increases the number of vertices and the number of edges by $j$. Therefore, $\chi$ does not vary. If, on the contrary, at some toroidal position $j \ge 2$  filaments merge (Fig.~\ref{FIG:7}), the number of edges increases by $j$ and the number of vertices by 1. So the characteristic decreases by $j - 1$. 

When we complete a full toroidal excursion, the characteristic would decrease each time that some filaments merge. Let us assume that there is a decrease in the number of components in going from section $s$ to the next section $s + 1$, i.e. $b_0^{2D} \left( s \right) - b_0^{2D} \left( {s + 1} \right) > 0$. Then the characteristic $\chi$ will decrease by $b_0^{2D} \left( s \right) - b_0^{2D} \left( {s + 1} \right)$  in going from $s$ to $s + 1$.  Here, $b_0^{2D} \left( s \right)$ is the 2-D Betti number $b_0$ of the structures in the $s$ toroidal cut.  Since the system is periodic, we find in the last section the $E$ vertices already counted at the beginning. Therefore, the characteristic of Euler-Poincar\'e for the full 3-D structure is 

\begin{equation}
\label{eq:chi_cal}
\chi  =  - \sum\limits_{s = 1}^S {M_1 \left( s \right)},
\end{equation}
where $M_1(s)$ denotes the number of mergers between sections $s$ and $s + 1$, i.e.  $M_1 \left( s \right) = b_0^{2D} \left( s \right) - b_0^{2D} \left( {s + 1} \right)$ when this number is positive and 0 otherwise. $S$ is the total number of toroidal sections and $M_1(S)$ is the number of mergers between sections $S$ and 1.

In these calculations we are assuming that mergers and bifurcations do not occur simultaneously. This is true for the analytical models that we use provided that we take a sufficiently large number of toroidal sections.

In case of a single ballooning mode and for low values of $\Phi_0$, $b_0 = 1$ and $b_2 = 0$, therefore  $b_1  = 1 + \sum\nolimits_{s = 1}^S {M_1 \left( s \right)} $.  However, this is not true in general and when we use this method in more complicated structures, we will only make the determination of the characteristic of Euler-Poincar\'e.

\begin{figure}
\centering
\includegraphics[angle=0,width=15cm]{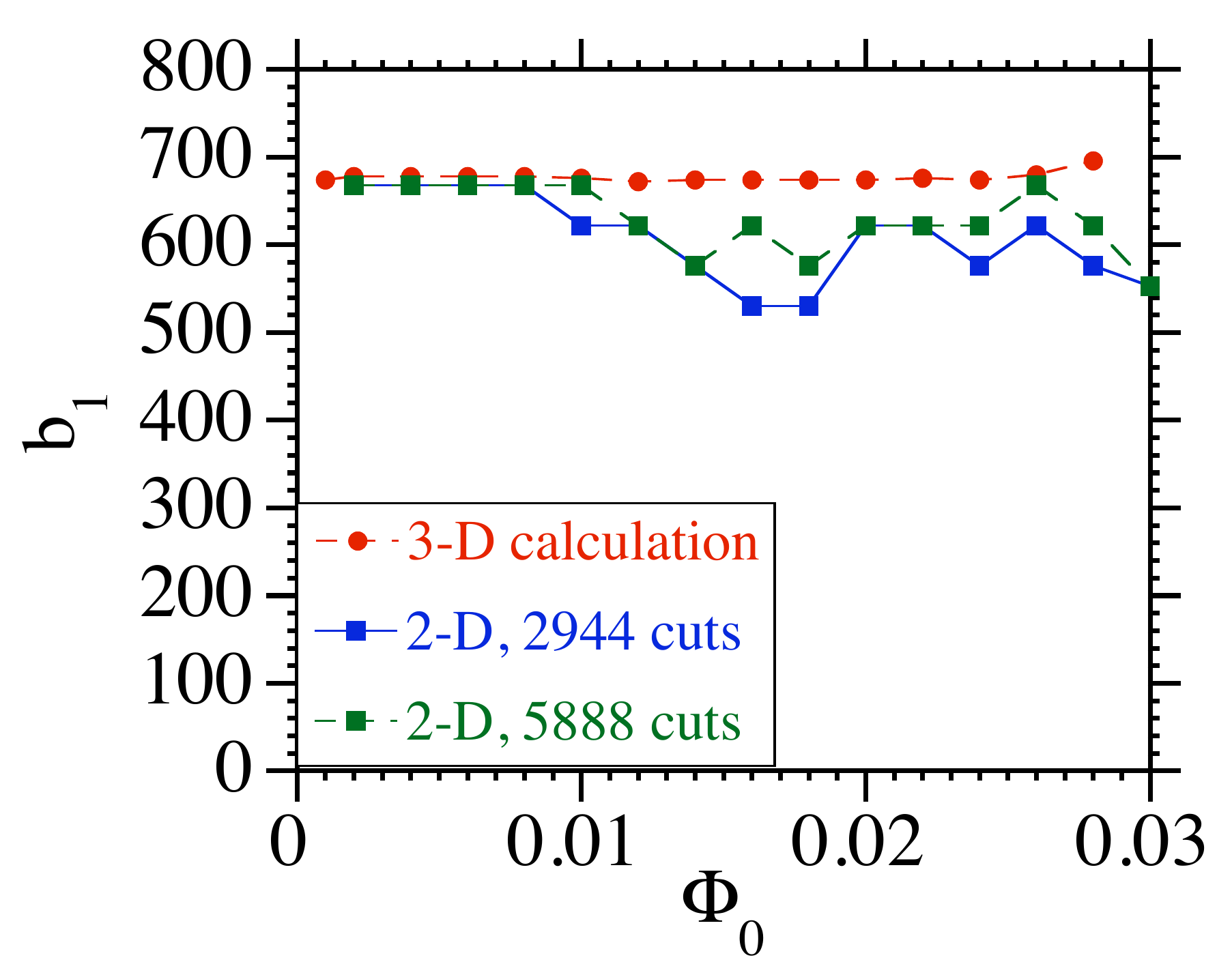}
\caption{Comparison of the results for the Betti number $b_1$ as a
function of $\Phi_0$ calculated with the methods of Section
\ref{sec:calculation} and Section \ref{sec:alternative}. For the
first, using $N_\rho=N_\theta=2N_\zeta=1200$, and for the second using
a 2-D Cartesian grid of $8000 \times 8000$. The latter has been done
for different number of toroidal cuts.}
\label{FIG:8}
\end{figure}

As pointed out above, the main problem with this approach is that, if we do not use enough toroidal sections, we might miss some of the filaments mergers. Therefore, this type of calculation can give us a lower bound to the Betti numbers. As an example, we calculate the $b_1$ number for the same $N = 23$ structure as in Fig.~\ref{FIG:5} using the direct computation explained in Section \ref{sec:calculation} and the method just described.  In Fig.~\ref{FIG:8}, we compare the results for both methods. We have plotted $b_1$ calculated with the method of Section \ref{sec:calculation}  in the case of highest resolution, $N_\rho =  N_\theta  =  2N_\zeta=1200$, as a function of $\Phi_0$ for a range of values in which there is not much variation on the results. We have also plotted the results given by the method discussed in this section for a 2-D Cartesian grid of $8000 \times 8000$ and using different numbers of toroidal cuts. We can see a relatively good agreement between the two calculations, although a very large number of toroidal cuts are needed to get converged results.

For a single toroidal mode the approach of this section can work very well. However, when we have a broad spectrum of toroidal modes the situation is more complicated, essentially due to the fact that we can have loops in the 2-D cuts. The break of a 2-D loop does not change the Euler-Poincar\'e characteristic. The merger of two portions of a connected structure to form a new loop decreases $\chi$ by 1. In case that a loop contract to a point, the characteristic increases by 1, but this phenomenon has never been observed in the structures we are studying and we are not going to take it into account.  Therefore now we should calculate the characteristic as follows. Let us assume that we consider $S$ toroidal cuts and denote by  $b_0^{2D} \left( s \right)$ and $b_1^{2D} \left( s \right)$ the Betti numbers of the 2-D structures in the $s$ toroidal cut. We define

\begin{equation}
\label{eq:mergers1}
M_1 \left( s \right) = \left\{ {\begin{array}{*{20}c}
   {b_0^{2D} \left( s \right) - b_0^{2D} \left( {s + 1} \right)\quad {\text{if}}\quad b_0^{2D} \left( s \right) - b_0^{2D} \left( {s + 1} \right) > 0}  \\
   {\;\;\;\;0\quad \quad \quad \quad \quad \quad \quad \quad \quad \;{\text{if}}\quad b_0^{2D} \left( s \right) - b_0^{2D} \left( {s + 1} \right) \le 0\quad }  \\

 \end{array} } \right.
\end{equation}
\begin{equation}
\label{eq:mergers2}
M_2 \left( s \right) = \left\{ {\begin{array}{*{20}c}
   {b_1^{2D} \left( {s + 1} \right) - b_1^{2D} \left( s \right)\quad {\text{if}}\quad b_1^{2D} \left( s \right) - b_1^{2D} \left( {s + 1} \right) < 0}  \\
   {\;\;\;\;0\quad \quad \quad \quad \quad \quad \quad \quad \quad \; {\text{if}}\quad b_1^{2D} \left( s \right) - b_1^{2D} \left( {s + 1} \right) \ge 0\quad }  \\

 \end{array} } \right.
\end{equation}

The characteristic of Euler-Poincar\'e of the 3-D structure is
\begin{equation}
\label{eq:chi_cal1}
\chi  =  - \sum\limits_{s = 1}^S {\left[ {M_1 \left( s \right) + M_2 \left( s \right)} \right]}.
\end{equation}

This is the expression we use for the more general nonlinear flow structures that we find in the 3-D nonlinear calculations.

\section{AN ANALYTICAL MODEL FOR THE BETTI NUMBERS OF THE BALLOONING MODES}
\label{sec:analytical}

Let us calculate the Betti numbers for a ballooning mode with the structure given by Eq.~(\ref{eq:single_mode}). We will use the approach described in the previous section to do the calculation of the Betti numbers.

One way of understanding the structures in 2ÐD given by Eq.~(\ref{eq:single_mode}) for constant values of $\zeta$ is by first making a model for the different components of the ballooning mode. We identify each component $m$ with the position of its $m$ maxima. They are located at the rational surface $m/N$ and at the poloidal positions

\begin{equation}
\label{eq:max_polpos}
\theta _k  = \frac{{4k + 1}}
{m}\frac{\pi }{2} - N\frac{\zeta }{m},\quad  - \frac{m}{2} \le k \le \frac{{m - 1}}{2}.
\end{equation}

Let us consider a field period,  $\pi / \left( {2N} \right) \le \zeta  \le 5\pi / \left( {2N} \right)$. We begin the field period at $\pi / \left( {2N} \right)$ because at this toroidal cut the distribution of structures has maximum symmetry.

\begin{figure}
\centering
\includegraphics[angle=0,width=7cm]{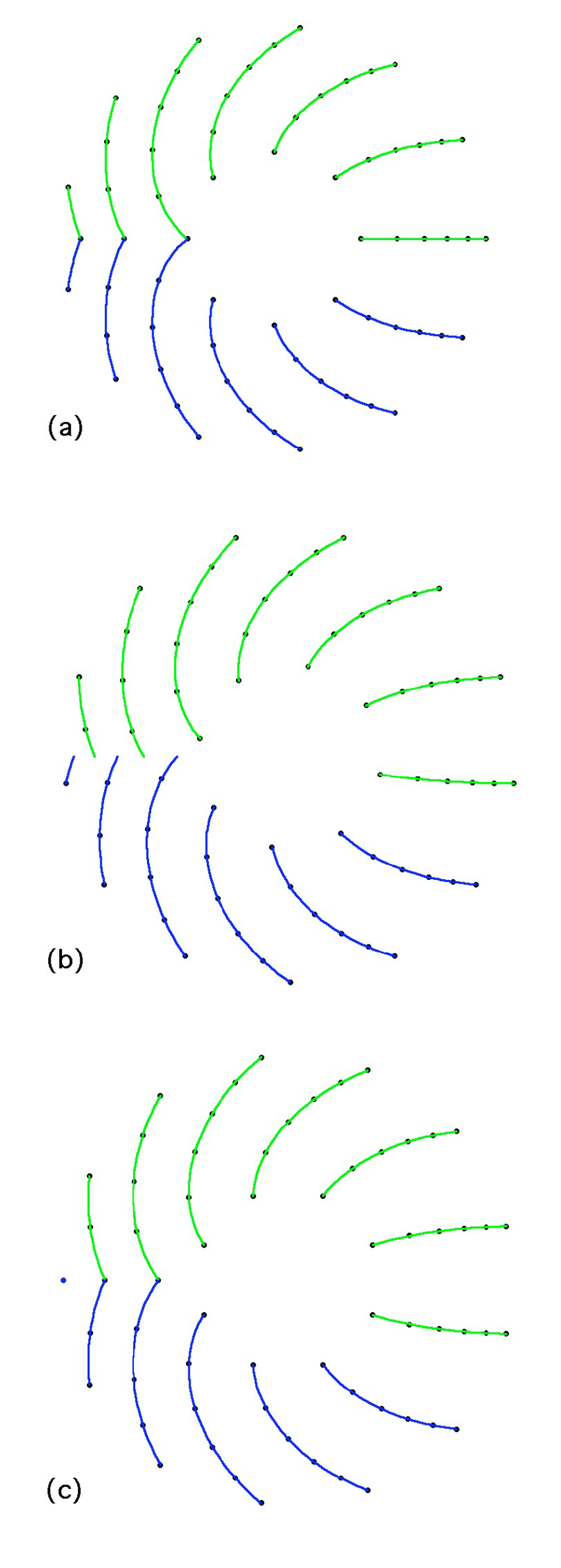}
\caption{Structures from the merger of the maxima of $\Phi$ at
different radii for the same $N = 7$ flow structure as Figure
\ref{FIG:6} for the same three toroidal cuts shown in that
figure. Note that the lines joining the maxima correspond to the black
structures in Figure \ref{FIG:6}.}
\label{FIG:9}
\end{figure}

For simplicity we assume that $\Phi$ has been normalized in a way that its maximum value is 1. We consider a range of values for $\Phi_0$ ranging from 0.001 to about 0.01. In the next section, we will discuss in more detail the range of validity. The regions of local maxima of $\Phi$ tend to merge forming streamers. This merger can be realized by joining the local maxima at different radii without crossing lines of minima. For $N = 7$, and in Fig.~\ref{FIG:9}, we show the resulting structures from the merger of the maxima. Note that the lines joining the maxima correspond to the black structures in Fig. ~\ref{FIG:6}.

In Eq.~(\ref{eq:single_mode}), $m_0$ and $m_1$ are the lower and upper limits of the sum over poloidal modes. Assuming that $m_0$ is even and $m_1$ is odd and using the model for the structures represented in Fig.~\ref{FIG:9}, we count the number of structures in this toroidal cut. For very low $\Phi_0$, we have the following structures:

\newcounter{struct}
\begin{list}{ \arabic{struct})}{\usecounter{struct}
	\setlength{\labelwidth}{0.75cm}\setlength{\leftmargin}{1cm}
	\setlength{\labelsep}{0.2cm}\setlength{\rightmargin}{0cm}
	\setlength{\parsep}{0.5ex plus0.2ex minus0.1ex}
	\setlength{\itemsep}{0ex plus0.2ex} }
	
\item The one at $\theta=0$, which is common to all.

\item The $(m_0-2)/2$ pairs before the merge of the positive and negative branches.

\item The $(m_1-m_0+1)/2$ structures that merge across the $x$-axis.

\end{list}
The total is
\begin{equation}
\label{eq:number_struct}
N_{S1}  = \frac{m_0+m_1-1}{2}.
\end{equation}

In other toroidal cuts, the number of structures varies due to bifurcations and mergers between structures. Let us look closely at these phenomena for that simple model.

The process of bifurcation and merger of the structures seems to follow a general pattern. All the $(m_1-m_0+1)/2$ structures that result from mergers across the $x$-axis on the left-hand-side bifurcate after a toroidal rotation of $\pi/(2N)$. Therefore, the number of components at this toroidal angle reaches its maximum value $m_1$. After another rotation of $\pi/(2N)$, we see that some of the bifurcated structures merge. In merging, they do not join the same structure as before, but the next one. This allows the rotation of the whole system. In the second half of the field period, the process reverses. After a field period is completed the structures are the same as the initial ones, but all maxima have performed a poloidal rotation of $2\pi/m$.

In short, the number of structures in each quarter of a field period is

\begin{equation}
\label{eq:quarter_struct}
\frac{{m_0  + m_1  - 1}}{2} \to m_1  \to \frac{{m_0  + m_1  + 1}}{2} \to m_1  \to \frac{{m_0  + m_1  - 1}}{2}.
\end{equation}

All bifurcations happen on the negative $x$-axis and so do mergers. The reason for this behavior is that the filaments in the structure move from left to right in the upper half plane and from the right in the lower half plane. Once a structure crosses the axis it is torn apart.  Here we have done the discussion of the structures for given parity of $m_0$ and $m_1$. For the general case, results are given in Table~\ref{TAB:I}.

For a low range of values of $\Phi_0$ and for the parameterization of Eq.~(\ref{eq:single_mode}), there is only one connected component in 3D and no voids. Therefore, we can use Eq.~(\ref{eq:mergers1}) to get $b_1$. Namely,
\begin{equation}
\label{eq:anal_betti}
b_0  = 1,\quad b_1  = 1 + N\left( {m_1  - m_0 } \right),\quad {\text{and}}\quad b_2  = 0.
\end{equation}

The different parity combinations listed in Table~\ref{TAB:I} give the same value for the $b_1$ Betti number.

\begin{figure}
\centering
\includegraphics[angle=0,width=15cm]{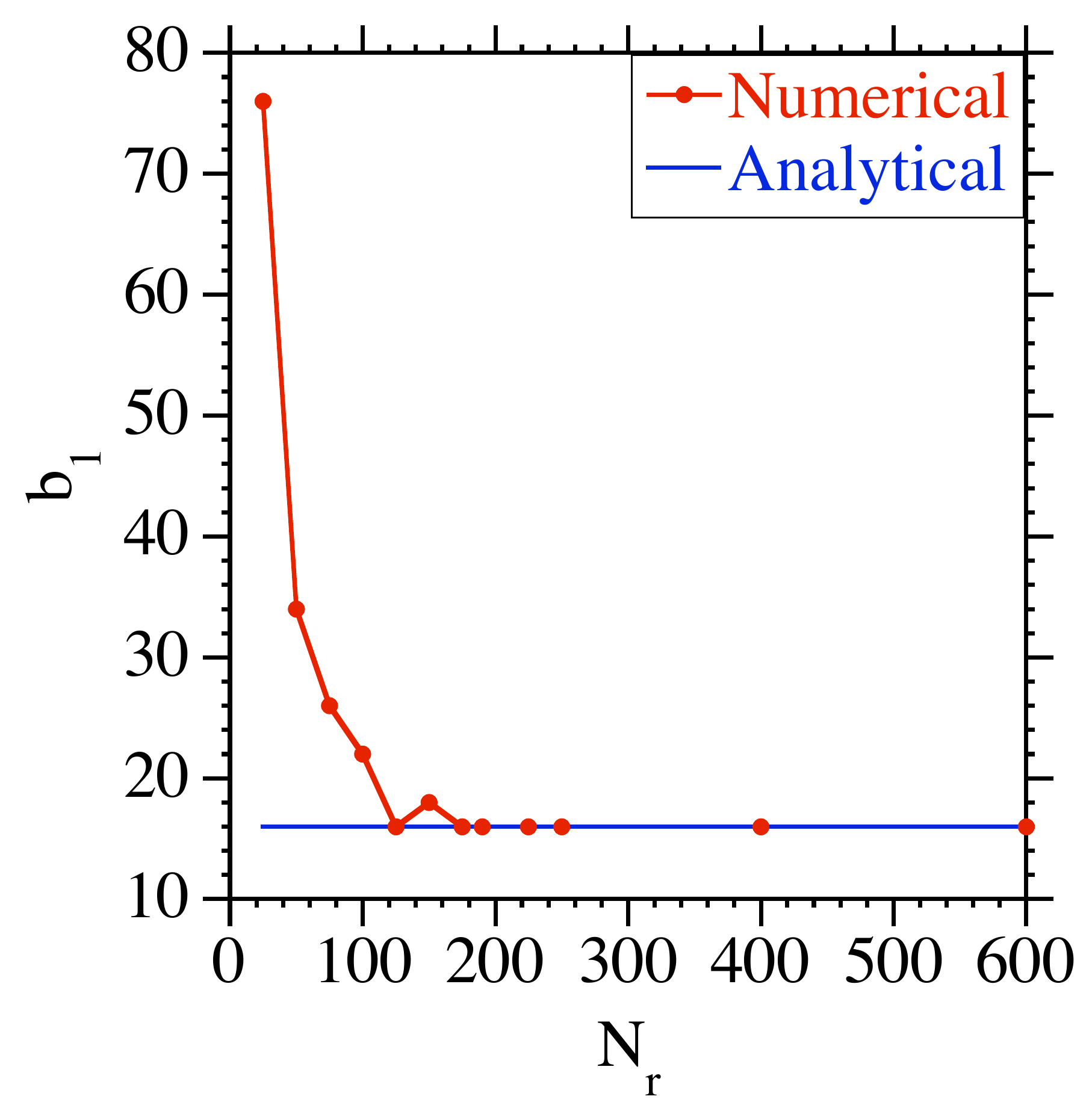}
\caption{Numerical results of the Betti number $b_1$ for a $N = 5$
flow structure converging to the value of the analytical model of
Section \ref{sec:analytical}, Eq.~(\ref{eq:anal_betti}), as the number
of cubes in the radial direction increases.}
\label{FIG:10}
\end{figure}

This (approximate) analytical calculation of the Betti numbers is very useful for the validation of numerical calculations. An example is shown in Fig.~\ref{FIG:10} for an $N = 5$ structure. We can see the numerical results converging to the value given by Eq.~(\ref{eq:anal_betti}) as the number of cubes in the radial direction increases.

For the numerical calculations in which the $NÊ=Ê23$ mode dominates, the relevant of poloidal mode numbers (modes such that $\Phi_m \ge \Phi_0$) is from $m_0Ê=Ê24$ to $m_1Ê=Ê53$. Using Eq.~(\ref{eq:anal_betti}), we obtain $b_1Ê=Ê668$, this value is consistent with the numerical results from the CHomP code discussed in Sect. \ref{sec:calculation}.

\section{SUMMARY AND CONCLUSIONS}
\label{sec:conclusions}

The results from Sections \ref{sec:calculation} and \ref{sec:alternative} show that we have a double approach in calculating the Betti numbers of a turbulent flow structure. One is by constructing a 3-D optimized covering of the structure with cubes. As the size of the cubes decreases and the number of cubes increases, the Betti number computations converge.

\begin{figure}
\centering
\includegraphics[angle=0,width=15cm]{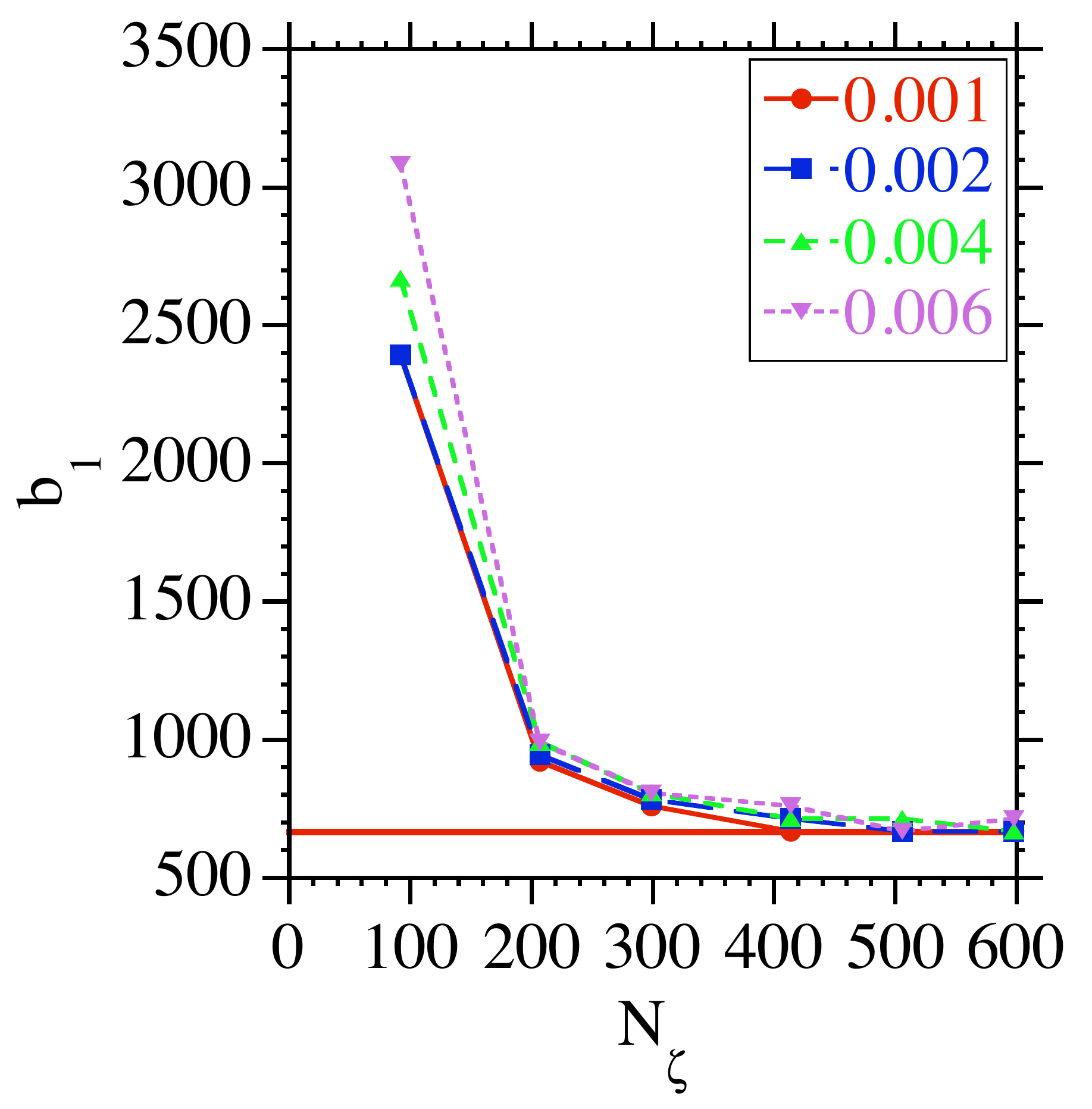}
\caption{The Betti number $b_1$ computed using the method of Section
\ref{sec:calculation} for a numerically calculated $N = 23$ flow
structure \cite{Garcia_02} and different values of
$\Phi_0/\Phi_{\rm{Max}}$. As the number of cubes increases $b_1$
converges from above. The horizontal line corresponds to the converged
value $b_1 = 668$.}
\label{FIG:11}
\end{figure}

\begin{figure}
\centering
\includegraphics[angle=0,width=15cm]{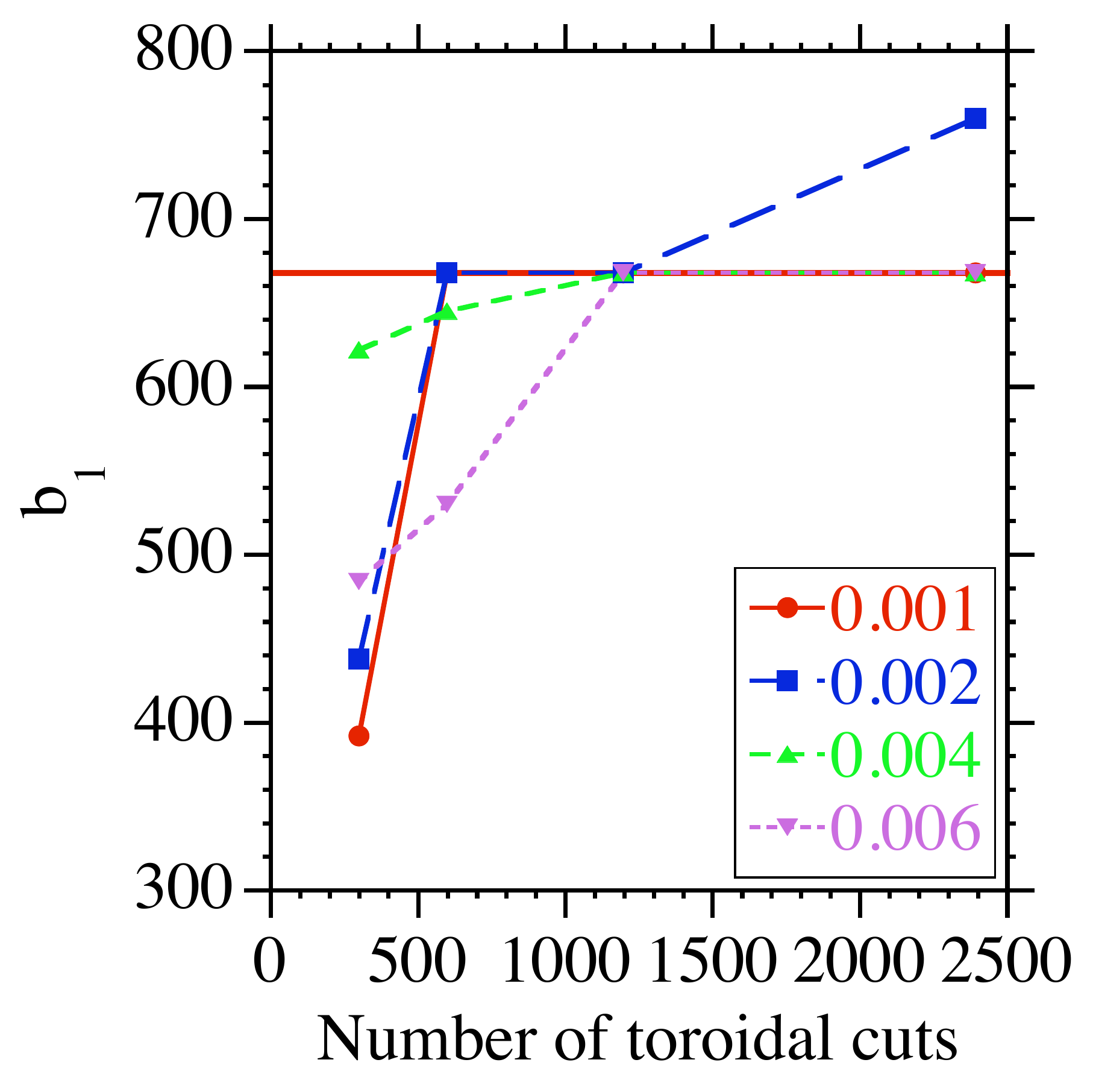}
\caption{The Betti number $b_1$ computed using the method of Section
\ref{sec:alternative} for the same cases as Fig. \ref{FIG:11}.  The
number of toroidal cuts is always a multiple of $N=23$, and
$N_\rho=N_\theta=2400$.}
\label{FIG:12}
\end{figure}

Consider the $N = 23$ structure of Ref.~\cite{Garcia_02}. The Betti number $b_1$ computed using the method of Section \ref{sec:calculation} is shown in Fig.~\ref{FIG:11} for different values of $\Phi_0/\Phi_{\rm{Max}}$. Here, $\Phi_{\rm{Max}}$ is the maximum value of $\Phi$ in the whole plasma volume. As the number of cubes increases $b_1$ converges from above. In these calculations, $N_\zeta$ is a multiple of $N$ and $N_\rho=N_\theta$ are multiple of 200 and close to $2N_\zeta$. The same calculation performed along the lines of Section \ref{sec:alternative} involves the determination of $b_1$ from a number of 2-D toroidal cuts through Eqs.~(\ref{eq:mergers1}) and (\ref{eq:mergers2}). The results are shown in Fig.~\ref{FIG:12}. Here, $N_\rho=N_\theta=2400$, and the number of toroidal cuts is always a multiple of $N$. As the number of toroidal cuts increases, $b_1$ converges from below as it is shown in Fig.~\ref{FIG:12}. The only exception is the case with $N_\zeta=2392$ and $\Phi_0/\Phi_{\rm{Max}}=0.002$, that could indicate that we need higher resolution in the $(\rho,\theta)$ plane for that value of $N_\zeta$. This convergence behavior gives a nice way to estimate this Betti number in more complicated structures. By applying the two techniques for calculating $b_1$, we can have an upper and a lower bound to its real value.

\begin{figure}
\centering
\includegraphics[angle=0,width=15cm]{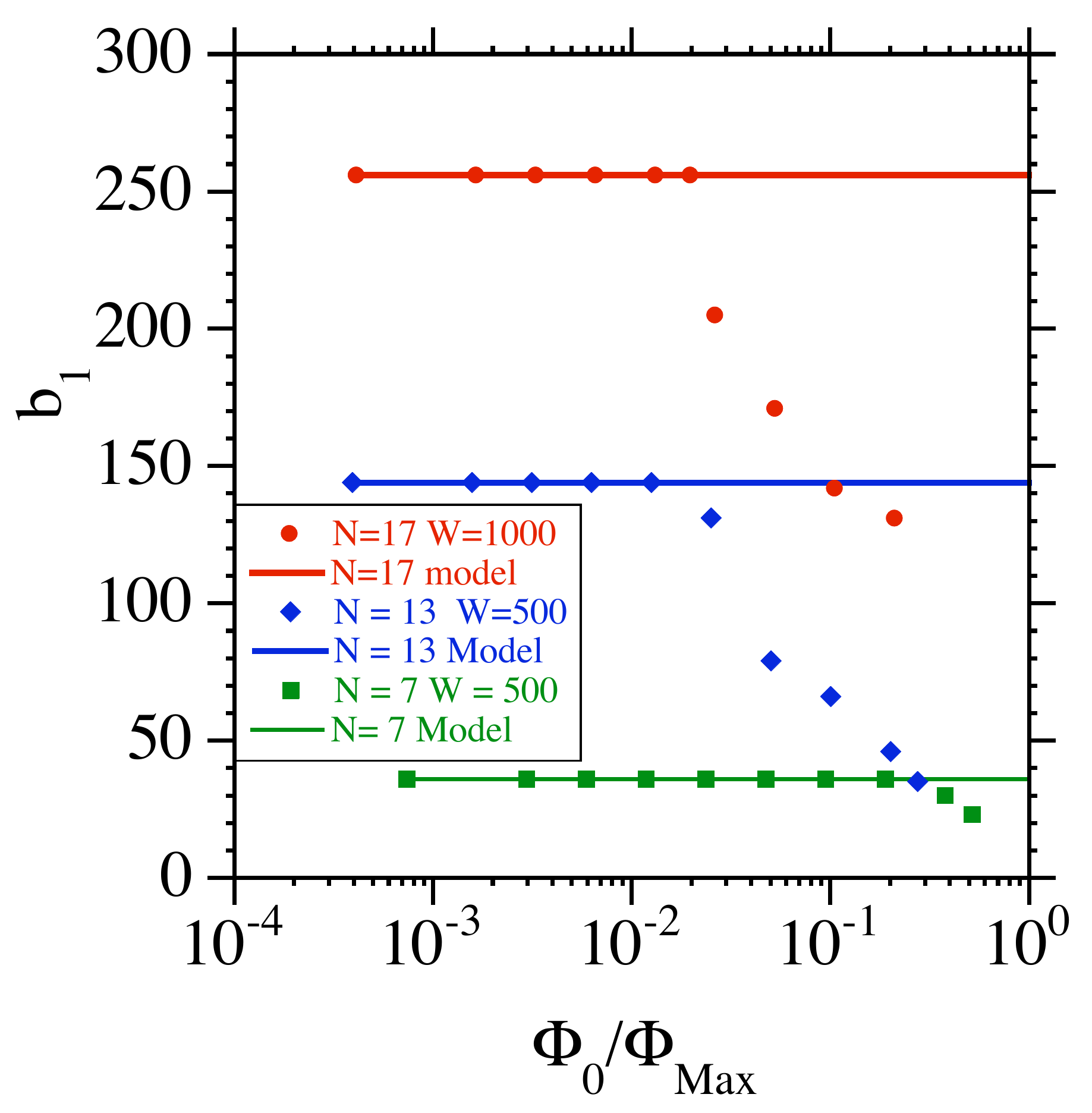}
\caption{The Betti number $b_1$ as a function of
$\Phi_0/\Phi_{\rm{Max}}$ for three different values of $N$. It shows
the range of values of $\Phi_0$ in which $b_1$ is constant. The figure
also shows that the value of $b_1$ in this range of $\Phi_0$ is the
one given by the analytical calculation of Section
\ref{sec:analytical}.}
\label{FIG:13}
\end{figure}

The relevant range of values of $\Phi_0$ is another important consequence of these studies. It seems that the optimal range is $0 < \Phi_0/\Phi_{\rm{Max}} \le 0.01$. The Betti numbers are practically constant in this range. In Fig.~\ref{FIG:13}, we show the Betti number $b_1$ as a function of $\Phi_0/\Phi_{\rm{Max}}$ for three different values of $N$. Note that $b_1$ is constant and has its maximum value in this low range of values of $\Phi_0$. The figure also shows that the value of $b_1$ in this range of $\Phi_0$ is the one given by the analytical calculation of Section \ref{sec:analytical}. For all these reasons, the choice of this range for $\Phi_0$ seems to be the best to characterize the structure. As $\Phi_0$ increases, the flow filaments become narrower and break, causing a drop of $b_1$. 

In conclusion, we have shown that it is possible to give a quantitative characterization of the turbulent plasma flows by determining the Betti number corresponding to structures of constant values of the stream function. The CHomP project offers a broad set of tools to carry out the numerical calculations.

Having a good resolution of the surfaces is essential. The intricate topological structures of the ballooning modes can lead to many false values for the number of loops and careful convergence studies are required.

We have used two approaches in those calculations:

\newcounter{approach}
\begin{list}{ \arabic{approach})}{\usecounter{approach}
	\setlength{\labelwidth}{0.75cm}\setlength{\leftmargin}{1cm}
	\setlength{\labelsep}{0.2cm}\setlength{\rightmargin}{0cm}
	\setlength{\parsep}{0.5ex plus0.2ex minus0.1ex}
	\setlength{\itemsep}{0ex plus0.2ex} }
	
\item Constructing a full 3-D file of cubes describing the structure.  We have found a way of getting optimal resolution for the determination of the cubes. This approach tends to give an upper bound of the 3-D $b_1$ Betti number.

\item Using multiple toroidal cuts and determining the Euler Poincar\'e characteristic from the number of mergers of 2-D components. This approach gives a lower bound to the 3-D $b_1$ Betti number.

\end{list}

For a simple fixed $N$ ballooning modes, the two approaches give in general the same value. For more complicated structures, we will use the two approaches in order to get upper and lower bounds for the Betti numbers.

The next step in this research is to characterize the variation, in time and $\beta$, of the Betti numbers for the flows obtained in 3-D resistive pressure-gradient-driven turbulence calculations.

\begin{acknowledgments}
We are grateful to the CHomP group for providing us with the software that we needed to carry out the calculations presented here. In particular, we acknowledge very useful discussions with Marcio Gameiro, Pawel Pilarczyk and Konstantin Mischaikow from the CHomP group.  Ones of us (BAC) is grateful to the Asociaci\'on EURATOM-CIEMAT for providing travel expenses. Part of this work is supported by the DGI (Direcci\'on General de Investigaci\'on) of Spain under project No. ENE2006-15244-C03-01 and by CM-UC3M (Comunidad de Madrid - Universidad Carlos III) Project No. CCG07-UC3M/ESP-3407.
\end{acknowledgments}

\clearpage

\begin{table}[]
\centering

\caption{Number of 2-D connected components at different toroidal sections.}
\vspace{0.2cm}
\begin{tabular}{lccccc}

  \hline
  \hline
   & \hspace{0.25cm} $\pi/(2N)$ \hspace{0.25cm} & \hspace{0.25cm} $\pi/N$ \hspace{0.25cm} & \hspace{0.25cm} $3\pi/(2N)$ \hspace{0.25cm} & \hspace{0.25cm} $2\pi/N$ \hspace{0.25cm} & \hspace{0.25cm} $5\pi/(2N)$ \hspace{0.25cm} \\
  \hline
 $m_0$ even, $m_1$ even \hspace{0.2cm} & $\frac{m_0+m_1}{2}$ & $m_1$ & $\frac{m_0+m_1}{2}$ & $m_1$ & $\frac{m_0+m_1}{2}$ \\
  $m_0$ odd, $m_1$ odd& $\frac{m_0+m_1}{2}$ & $m_1$ & $\frac{m_0+m_1}{2}$ & $m_1$ & $\frac{m_0+m_1}{2}$ \\
  $m_0$ even, $m_1$ odd& $\frac{m_0+m_1-1}{2}$ & $m_1$ & $\frac{m_0+m_1+1}{2}$ & $m_1$ & $\frac{m_0+m_1-1}{2}$ \\
 $m_0$ odd, $m_1$ even& $\frac{m_0+m_1+1}{2}$ & $m_1$ & $\frac{m_0+m_1-1}{2}$ & $m_1$ & $\frac{m_0+m_1+1}{2}$ \\
  \hline
  \hline

\end{tabular}
\label{TAB:I}
\end{table}

\clearpage

\newpage

\newpage

\newpage

\newpage

\newpage

\newpage

\newpage

\newpage

\newpage

\newpage

\newpage

\newpage

\end{document}